\documentclass[journal]{IEEEtran}
\usepackage{xcolor,soul,framed} 

\colorlet{shadecolor}{yellow}
\usepackage[pdftex]{graphicx}
\graphicspath{{../pdf/}{../jpeg/}}
\DeclareGraphicsExtensions{.pdf,.jpeg,.png}

\usepackage[cmex10]{amsmath}
\usepackage{array}
\usepackage{mdwmath}
\usepackage{mdwtab}
\usepackage{eqparbox}
\usepackage{url}

\begin{document}
\bstctlcite{IEEEexample:BSTcontrol}
    \title{A Survey on Limitation, Security and Privacy Issues on Additive Manufacturing}
  \author{Md~Nazmul~Islam,~\IEEEmembership{}
  Yazhou~Tu,~\IEEEmembership{Member, IEEE,}
  Md Imran Hossen,~\IEEEmembership{}
  Shengmin Guo, and~\IEEEmembership{}
  Xiali~Hei,~\IEEEmembership{Senior Member,~IEEE}\\

}



\maketitle

\begin{abstract}
Additive manufacturing (AM) is growing as fast as anyone can imagine, and it is now a multi-billion-dollar industry. AM becomes popular in a variety of sectors, such as automotive, aerospace, biomedical, and pharmaceutical, for producing parts/ components/ subsystems. However, current AM technologies can face vast risks of security issues and privacy loss. For the security of AM process, many researchers are working on the defense mechanism to countermeasure such security concerns and finding efficient ways to eliminate those risks. Researchers have also been conducting experiments to establish a secure framework for the user’s privacy and security components. This survey consists of four sections. In the first section, we will explore the relevant limitations of additive manufacturing in terms of printing capability, security, and possible solutions. The second section will present different kinds of attacks on AM and their effects. The next part will analyze and discuss the mechanisms and frameworks for access control and authentication for AM devices. The final section examines the security issues in various industrial sectors and provides the observations on the security of the additive manufacturing process.
\end{abstract}


%
\IEEEpeerreviewmaketitle


\section{Introduction} 

\IEEEPARstart{A}{dditive Manufacturing (AM)} is mainly known as the technology that builds 3D objects by adding layer-upon-layer of material including plastic, concrete, metal, or by far extent bio materials. Recently, for AM, several machines are available to produce industry-quality products. Those machines are typically controlled by advanced sensors, and they need very high processing power. AM has several advantages over traditional manufacturing. It requires a shorter time to produce industrial components with very complex structure rather than the conventional process. Production of industrial functional parts with less material waste and optimized physical structures of the products made AM very much popular. Thus, the AM industry is becoming a multi-billion-dollar industry. 

As AM increasingly becoming popular day by day for its shorter production time, better durability factor, and better working efficiency, the global market of the AM industry are on the rise. The global additive manufacturing market study~\cite{jiang2017predicting} found that 18.6 $\%$ of growth in production will produce in the 2015-2025 period. In 2014, Bourell et al.~\cite{bourell2014roadmap} presented a workshop paper about the roadmap and the impact of additive manufacturing. In 2017, another report from Arizton Advisory $\&$ Intelligence’s~\cite{miller2019investigating} predicted that the 3D printing industry would gain revenue of about $\$11$ billion by the end of the year 2022. Attaran et al. presented the idea of rapid prototyping and the main advantages of AM over traditional manufacturing~\cite{attaran2017rise}.  Ernst and Young conducted a study, which shows the rapid adaptation of additive manufacturing worldwide~\cite{muller2016will}. In the United States alone, about 18\% of companies are now connected to additive manufacturing implicitly or explicitly, and that number will go up in the coming years.

With the increasing dominance of the AM industry, this technology with significant capabilities and reach of production also has several potential misuses that can cause significant concerns for the users. As AM is adopting rapidly in manufacturing industries and now producing multi-billion-dollar revenue, these can motivate adversaries to perform a more extensive array in cyber- and cyber-physical attacks. Manufacturers are yet to implement a robust security system for additive manufacturing devices. So, experts have given warnings about the potential security risks of a large number of AM devices that connected to public networks, and the technical feasibility of cyber-physical attacks. Belikovetsky et al. conducted a study called dr0wned that sabotaged a propeller design~\cite{belikovetsky2017dr0wned}. They showed that a 3D-printed sabotaged propeller broke in the mid-air after a short flight time causing a fall and suffered significant damage.

Privacy and security remain a significant issue for additive manufacturing, which poses a great concern for industrial manufactures and the end-users. So, the manufacturers and the end-users must understand corresponding cyber- and cyber-physical potential attacks which include attacks relating to communication medium and manufacturing process. The primary aim of this survey is to educate manufacturers and the end-users about potential security concerns and offenses related to additive manufacturing. Also, for the researchers who are just entering research related to AM security, we provide an in-depth discussion on this multi-disciplinary research field.  

There are many published surveys on additive manufacturing related to production and production effects in industrial sectors. There are several surveys on additive manufacturing in which the authors enlisted several security concerns on AM. Tang et al. surveyed design methods for additive manufacturing to improve functional performance ~\cite{tang2016survey}. They showed how significant is the design change for the AM to improve the performance of any functional parts. They introduced the differences between general lattice and homogeneous lattice structure to create an effective design for the 3D printed medical devices which performed way better than traditional manufacturing parts. Yampolskiy et al. focused on the different attack taxonomy and security measures in their survey. They proposed a basic framework for the 3D-printing, and they categorized the affected parts of the manufacturing devices in their survey~\cite{yampolskiy2018security}.

In this survey paper, we identified and explored the additive manufacturing constraint, security, and privacy issues. The first part of this survey presents the most critical limitations of additive manufacturing and their probable solutions. In the following section, we discussed the classifications of the existing attacking strategies on AM and their effects. Then we explored the mechanism and authentication for the access control of AM devices in the third part. Finally, we analyzed security issues in various industrial sectors for the AM and conclude the survey with a summary of our observations and identified gaps.

\section{AM LIMITATIONS $\&$ POSSIBLE SOLUTIONS} 
Why is it challenging to implement robust security architecture and apply security features to additive manufacturing? Mhapsekar et al. presented two constraints~\cite{mhapsekar2018additive}. Minimization of several thin layers is the first limitation. Minimization of the volume of support structures is the second limitation. The main limitations are due to machine capabilities, process characteristics, general material, and process specific constraints, AM benchmark constraints, constraints related to material properties and processing, limitations relating to quality control and external restraint. Additionally, in AM, which known as 3D-printing, there are some severe malicious implications for the software and the desktop 3D firmware. 

\subsection{Constraints Due to Machine Capabilities and Process Characteristics}
In general, every additively manufactured parts is affected by the process family and the device used to build the components. Process specification includes the general material deposition method, the principle of bonding, methods of re-coating. These determine the types (i.e., polymer, concrete, etc.) and nature (i.e., solid, powder, etc.) of the raw materials. Machine-specific capabilities and limitations also include several materials. At the industry level, a particular machine can only operate on unique contents. The minimum build resolution (generally in x, y, and z-axis) and build dimension (also in x, y, and z-axis) can vary from material to material. Process-wise faults like unintentional internal voids, blind holes can present due to wrong material choice. This problem can be solved by monitoring the post-processing level and quit if there any faults showing in manufacturing levels. These types of fundamental limitations could be removed by building the parts in another machine.

\subsection{General Material $\&$ Process Specification Constraints}
Since the revolution of the AM in industrial sectors, multiple manufacturing, and production design guides have published. Those guides typically focused on process specification guidelines, machine constraints, and considerations. In 2015, materialise~\cite{materialise2015design} published 19 design guides for a variety of materials. Each design provides precise information about the materials like minimum wall thickness, minimum detail size, maximum part size, expected accuracy, and those considered as ‘design specification.’ Stratasys published three guides mentioned DMLS~\cite{thompson2016design}, FDM~\cite{thompson2016design} and laser sintering, which are sets of ‘basic rules, tips, and tricks’ for process and material specific. Each guide includes several general specifications like supported and unsupported wire thickness, wire size, maximum and minimum bounding box, a minimum size of the escape hole for entrapped material, the expected outlook of the material, and the expected accuracy.

Lastly, 3D systems published two design guides, both focusing application-specific considerations for brass~\cite{bernard2019functional} and plastic SLS components~\cite{thompson2016design}. Those two guides include features like internal channels, assemblies, springs, hinges, threads, and snap fits. Adam and Zimmer presented a catalog of design rules that include laser melting, laser sintering, and FDM, which addressed geometric constraints like element transitions, sharp edges, feature spacing, and unsupported features~\cite{adam2014design}.

\subsection{AM Benchmark Constraint}
Learning design rules/guidelines is an advantageous starting point. 
But they do not provide general information about individual machines and their capabilities. So, when the detailed information is required to support the design and manufacture products, a benchmark is the most crucial factor to study and compare AM processes. In the beginning, AM benchmarks introduced for comparison, selection, and process optimization. They were relatively large features and can quickly characterize. With the progress of design and processes, AM benchmarks gained more ‘specific’ characteristics (i.e., holes, wall thickness, angles, notches, bosses, towers, excellent features, etc.~\cite{moylan2012review}).

\subsection{Constraints Related to Material Properties $\&$ Processing}
For AMs, in many cases, raw materials can be used without any modification, and some elements need to be adopted before they apply for manufacturing production parts. Some materials and design specifications need to be changed at every stage of production with the manufacturing needs change from stage to stage. Like, laser sintering gold requires a change in the alloy to prevent material evaporating~\cite{singla20163d}. Also, the material properties of the final parts can be a part of modification in AM processing. Material properties affect the use of recycled raw materials and the recycling process. So, the potential degradation in quality associated with the cost and waste of the manufacturing materials. 

\subsection{Constraints Related to Quality Control $\&$ External Constraints}
Though AM brings an excellent prospect for the industrial production for the design purposes, it has a significant issue for the quality control beyond the output. Verification of materials, geometries, and surfaces is one of the problems related to post-product quality control. It is very challenging when AM creates geometry and the part material at the same time. At the end of the printing process, it is urgent to check undesirable characteristics, unintentional internal voids, and defected porosity in the bulk material. For the functional content, those challenges go higher, and it is required to fix the optical and mechanical property characteristics before the product goes for marketing.  

Additive manufacturing requires complex design and a very high level of data processing. For those, there have been several challenges for the entire process. The first and most crucial challenge about the present specification system, which defined in the ISO~\cite{feeney2015portrait}, is that AM could not cover complex freedom shapes. There also have some requirements for the communication system~\cite{kiontke2015freeform} at the time of manufacturing. There does not exist much significant work in this area until this time. The manufacturing designers need to keep in mind that the first stage choices of the design have a substantial impact on the production and the quality control of the manufacturing process. So, the basic design related to quality control and metrology is a part of AM.

AM can be handy for the medical and aerospace industry. But, these two industries are highly regulated, and every newly manufactured product always require regulatory approval. So, it is a vital and urgent need for the designers to test, verify, and document all the process of manufacturing to get regulatory approval. As AM industry does not have much historical data, obtaining regulatory approval is one of the significant challenges. Currently, AM primarily applied in the military sector for the betterment of military aircraft than commercial  aircraft~\cite{campbell2012additive}. Parts related to commercial aviation is very sensitive compare than military aircraft. Without safety testing and regulatory approval, the AM industry can't manufacture new pieces for the business jets.   

In the medical sector, there has been a very significant improvement for AM. Recently, researchers from the Tel Aviv University of Israel created a 3D-printed heart model, and they printed that model with blood vessels using patient's cells. That was the very first 3D-printed human organ. There have been a significant number of medical devices printed using 3D printing technology, and they are now providing service in several medical institutions all over the world. If a patient permits to perform any experiment for medical purposes that should name as `single-use experimental' basis experiment and that permission is called `explicit medical consent'. Those uses are limited in numbers compare to other medical device uses. Though, a large number of hip implants gained the approval of the regulatory committee in recent years.

\subsection{Limitations Due to Malicious Firmware}
As AM is rapidly growing and becoming very popular day by day, adversaries may try different ways to manipulate and endanger the manufacturing process. Researchers are always trying to find flaws to fix and improve the AM industry. From the beginning of designing the 3D model to the final print of that model, one crucial media that always needed to be maintained is printing software or otherwise known as firmware. In recent times, there have been several findings of severe complications of the 3D printer software. In 2016, Moore et al.~\cite{moore2016vulnerability} gave a vulnerability analysis of desktop 3D printer software. They analyzed a set of desktop 3D printer software to identify potential attack vectors. A basic set of analysis tools were used for their work. Their dynamic analysis of USB communication shows vulnerability in transferring packet data between desktop to a 3D printer, and for this monitoring, they used a device monitoring system. Their specific set of 3D printer software includes Cura 3D, ReplicatorG, Repetier-Host, and Marlin Firmware. They used VCG, RATS, Scitools, Fortify, and DMS analysis method to analyze that software. For Cura 3D software, RATS found occurrences of fixed buffers in global variables and use of uncontrolled strings. VCG found memory management issues (i.e., use of sscanf can be susceptible to buffer overflows). That issue can crash the system due to lack of resources. Scitools found a variety of coding style issues, and that includes code variable without comments and the dead code. But, not a direct effect that permits attacks but can hinder further developments. Log forging, path manipulation, command injection, unreleased resources, xml injection, location of dependent comparison, and null deference were found in ReplcatorG software. VCG reported some issues like unreleased resources, lack of exception handling, and getting the path from variable for ReplicatorG. Those issues can lead to file mishandling, integer overflow, and denial of service for the lack the resources. Files and functions that were too long or highly complex have some issues found by Scitools. For Repetier-Host software, VCG found three problems. They were improper thread locking, use-time/resource check-time race condition, comments in potential unfinished code. The authors found unreachable code, over-long functions, over-long files located in Repetier-Host using the analysis of Scitools. In Marlin firmware, the authors~\cite{moore2016vulnerability} found two issues (i.e., use of strcat and fixed global buffers). Incorrect use of those two can lead to a system crash. VCG discovered integer overflows for strlen and signed/unsigned comparison. Scitools reported dead code and variable without any comment.

The authors of~\cite{moore2016vulnerability} found another major flaw in dynamic communication between Repetier-Host and the 3D printer itself. 3D print always completes with the G-code command from the host. But, there are no authentication or encryption media used in transferring the G-code from the host to the printer. So, an adversary can quickly get the plain text of the G-code. The authors found the G-code could be modified at any end by the adversary to damage the final print, which can cause a significant hazard in the printer.   

In 2017, their follow-up work~\cite{moore2017implications} have shown the additive manufacturing industry about the implication of malicious firmware. They provide an example of "G-code packet capture" to capture format of the G-code, unmodified control loop, and process command function. They mentioned two categories of attack strategy and vulnerability. First one is the modification of the commands from the host, and the second one is the modification of the variable that will increase the extrude rate and create an unintentional hazard.

~\section{Attacks on AM and Their Defences}
Additive manufacturing is mainly a representation of computer-aided production ~\cite{gibson2014additive} and also can be called a Cyber-Physical System (CPS). Physical damage can occur by the attacks on CPS. So, the adversarial goal is to attack the cyber domain for the manipulation in the final output. 

~\subsection{Attack Vectors on AM}

In AMs, there are three categories of attack targets: the environment, the manufactured part, and the AM equipment~\cite{yampolskiy2016using}. An attack on AM could affect other equipment and machines nearby. If electrical components are compromised, then it could damage the whole device. Attack on the manufactured part could harm the final assembly. Embedding a defect in a rotating part at a predetermined speed has been already documented~\cite{belikovetsky2017dr0wned}. A manufactured part could be affected by an altercation in the process and in-process characteristics. Unexpected material could be introduced to the primary material to contaminate the manufacturing process and thus, the manufacturing object. 

AM equipment itself can be a target of attack. Mechanical components like powder handling equipment, the extruder of a 3D printer, seals, re-coating system, and various other part are vulnerable to the physical attacks. Electrical components like sensors, motors, power supplier, interlocks play a critical role in the manufacturing process. Likewise, mechanical parts, attacks relating to electrical components of AM devices can cause irreparable damage.

~\subsection{Attack Methods on AM}

From an adversary viewpoint, additive manufacturing is one of the prime sectors to exploit in the present time. As the inclusion of AM technology becoming more and more popular over the traditional manufacturing process, it became a prime target for the security experts. In this portion of our survey, we will present several attack methods on the additive manufacturing process.

Xiao et al.~\cite{xiao2013security} first presented in a keynote session at XCon2013 about the effect of attacks on Desktop 3D-printer. In that session, Xiao Zi Hang gave the idea that 'printing results,' including the size of the print object, material strength, printing layers, and position of the components, might change through an attack. He mentioned about the modification in software or configuration, object description or the firmware, etc. With the change or rewrite the code in the RepRap3D printer's firmware converts the print bed or extruder temperature twice than the temperature set by the G-code in the main module.

Faruque et al. ~\cite{faruque2016acoustic} presented the first paper related to producing the sound of a stepper motor of a 3D printer. The authors showed how acoustic emanations could be 'tied-back' to the actual movement of the stepper motors. Based on this, researchers can reconstruct the 3D printed object. Based on the recorded sound, they were able to restore the 3D print model with the accuracy of 78.35$\%$, which which could cause damages to the intellectual property of authentic manufacturers.

In~\cite{al2016forensics}, the follow-up work of~\cite{faruque2016acoustic}, the authors focused on thermal side channel. They used an infrared camera to capture the image, and based on image analysis; they tried to identify individual 3D printing process activity, for example, nozzle movements. The goal of the paper was to reconstruct a fully 3D printed model. They used the video feed to estimate test-bed motions by using an algorithm which they developed. They also used a mapping algorithm to transform the images into nozzle and base plate activity. The camera used for capturing the video feed was a low quality (50 Hz, non-auto focus, 640 x 480 resolution) single, fixed viewpoint. So, they assumed that a camera with higher quality would bring improvement in the accuracy of the algorithm.

In 2016, a side-channel attack against a wide variety of additive manufacturing components, which generally constructs and forms the object introduced by Hojjati et al.~\cite{hojjati2016leave}. Acoustic and magnetic side channels used in the attack were recorded by a compromised cell phone. After the analysis of the recording, the authors found that the tape can reconstruct the movements of the manufacturing equipment. For the line segment length, the reconstruction is accurate within one millimeter and one degree for the turning angle of the machine. The authors proposed an audio obfuscation or audio confusion method to block this type of attack. In the counter of the attack, they play the audio recording to other slightly changed manufacturing sessions. The loss of the harmonic is not the way to defend against this type of attack as the adversary can easily track the damage and find out the real recording.  

Song et al.~\cite{song2016my} proposed an attacking strategy that is similar to Faruque et al.~\cite{faruque2016acoustic} in terms of acoustic emission. But there are several differences between those works. Firstly, in~\cite{faruque2016acoustic}, they used a professional audio recorder to record the audio. But, in~\cite{song2016my}, the authors used sensors available on smartphones. Secondly, instead of using a single side channel, the of~\cite{song2016my} used magnetic and acoustic emanations. This approach has a more precise result rather than the previous one. The reconstruction model has only 5.87$\%$ error with this approach.

Gupta et al.~\cite{gupta2018vivo} demonstrated anti-counterfeit system based on compensatory manufacturing conditions and pairing defecting model features. Printer configured without material removal of embedded model results in a void or filled with support material in the final output. Such voids can create a defect and result in fracture under strain.  

Yampolskiy et al.~\cite{yampolskiy2014intellectual} identify that Internet Protocol or IP is not limited to 3D printed object geometry alone. 3D object geometry also requires manufacturing parameters and properties like operational parameters of a part. The authors also proposed an outsourcing model which provides service to the customers by the AM manufacturers and can depend on the IP manufacturing parameters. Third-party can give the IP manufacturing parameters. 

Do et al.~\cite{do2016data} analyzed two popular 3D printers; the MakerBot 3D printers and the Replicator (and the mini version of Replicator). Both the printers connected from the host via WiFi, USB, or Ethernet. In the analysis of the communication of the networks, the authors found the adversary on the local network can easily retrieve the previous 3D print models. So, they further investigated and found that user's authentication code and client secret are not encrypted. These credentials could be used to send separate commands from the host to the printers to perform unintentional jobs.  

Ilie et al.~\cite{ilie2017built} covered a lot about the failure for altering the parameters in the printing process. They showed that manipulation in the laser power and exposure time create layers with increased porosity in the building part. These layers become more vulnerable in the presence of stress. The authors showed deformation, and strategic failure for the manipulation. Above mentioned attacking strategy is the same as the sabotage attack.

Slaughter et al.~\cite{slaughter2017ensure} presented indirect sabotage on AMs. The authors showed that false reading for the sensors and the feedback loop in the existing process could be compromised. For AM metal with powder bed fusion (PBF), they showed an analysis about the attack could lead to the adjustment of the laser with false estimation and position of the laser either too high or
too low than the expected position for the printing. Incorrect
laser positioning and errors in laser strength lead to lack of porosity or
higher porosity in the final printed object. This lead to defects, and the printed object is vulnerable to mechanical stress.

~\subsection{Defense Measures for AM Sabotage}

Ding et al.~\cite{ding2015wire} published a paper in 2015, which presented the idea against the future attacking possibility and future interest in wire-feed additive manufacturing technique. 

Bayens et al.~\cite{bayens2017see} considered a threat model in which either the host PC or printer firmware compromised. When similar kind of objects manufactured multiple times, they proposed a three-layer framework for replication verification. So, the process does not depend on G-code. It actually depends on the replication of the object using acoustic, spectroscopic, and gyroscopic replication verification. They verified the system and proposed a model by printing a tibial knee implant. The environment of the printing process should be noise-free to have the perfect sound for the acoustic measurement.  They also mentioned the model could improve the performance using combined data acquisition as gyroscopes to have a higher degree of errors.

Chhetri et al.~\cite{chhetri2017cross} proposed a model for cross-domain attacks and cross-domain defense approaches for AM. The authors introduced Cross-domain security. Their proposed model covers a variety of security purposes for the manufacturing process. The model can work for: 1) determining information leakage, 2) abnormal behavior determination in the physical domain, 3) prediction of component and system failure, and 4) analysis of cross-domain attacks.

In 2015, Hou et al.~\cite{hou20153d} presented a 3D printing watermarking model that can survive the re-scanning process and 3D printing. A digital watermark is a kind of marker covertly embedded in a noise-tolerant signal such as audio, video, or image data. Each layer has an exterior in the x-y plane with the normal vector. An embedded watermarking model extracted with surface scanning  and a histogram of each layer could produce. Cho et al.~\cite{cho2006oblivious} performed several tests and was able to find out two printed models out of the four printed models by using the histogram analysis.  

Vincent et al.~\cite{vincent2015trojan} proposed a system that includes piezoelectric transducers (PZT) to measure the impedance and to distinguish between altered and unaltered parts to a manufactured part. The authors argue that the existing quality control (QC) system was not adequate to detect the sabotage attacks. That's why they introduced the PZT. Maintaining key quality characteristics (KQC) of a product generally is the best way to manage and control the quality of a product. The authors provided a scenario where modification of design file created an additional hole in the car body frame and thus reducing the strength of the mechanical property. The introduction of PZT showed a significant improvement in the context of a structural health monitoring system and for the work of Sturm et al.~\cite{sturm2016situ}.

Kennedy et al.~\cite{kennedy2017enhanced} proposed a solution for the counterfeit detection for the fused deposition modeling (FDM) 3D printing process. Every non-altered polylactic acid (PLA) used to print 3D part, while custom parts used to write a unique QR code. QR code generates a chemical signature, and the authors proposed to store these signatures as block-chain entries and searchable references.

Belikovetsky et al.~\cite{belikovetsky2017detecting} exploit acoustic side-channel to detect attacks on FDM desktop 3D printer which is similar kind of work performed by Bayens et al.~\cite{bayens2017see} and Chhetri et al.~\cite{chhetri2017cross}. The authors used an analysis process called Principal Component Analysis (PCA) to generate and compare the generated audio signal in a 3D printing process. The validity of the follow-up prints compared by using the signature of the prints against the genuine. They defined a set of activities like insertion, deletion, modification, or reordering in single G-code command. They determined a threshold in the process under which they considered a change in the process in a minimal second. This process is reliable to check abnormality like filament extrusion.

Sturm et al.~\cite{sturm2016situ} proposed a system in which they mentioned the coupled piezoelectric sensor influences part material. In a process, small fabrication can be easily measured using the sensor and for the impedance variance already affected layer can also be detected. Other prints verified for the production using the recording of the original signature or baseline signature. The authors claimed the detection of layer change with the help property called sensitivity. However, sensitivity could increase if the material density is high and decrease if the mass is lower. Also, it changes with the position of the sensors from the material at the time of production. 

Straub et al.~\cite{straub2017physical} proposed image-based fault detection and security in 3D printing. Their initial work used five different angles to track the printing process using the pixel image~\cite{straub2015initial}. Changes in camera position, background, angle changes are the most common factor for the degraded image. So, adapting the similar kind of way the authors went for defect detection, use of incorrect print materials and mispositioned prints. Rather than the initial computational analysis, the proposed method worked as practical applications rely on human operations in the manufacturing process. Print-Material detection is limited to changes in color and the reflectively of the object. 

DeSmit et al.~\cite{desmit2016cyber} proposed an intelligent manufacturing vulnerability assessment approach to work with the NIST framework~\cite{shackelford2015toward}. The manufacturing process is a set of entities that connects each item of the collection with the others. Human operations, cyber-physical design tools, physical resources, and equipment assessed adequately in process activity. The authors claimed that most of the vulnerabilities have occurred on the connections of those entities. The assessment set some self-directed questions for the manufacturer in terms of vulnerability measurement and labeling vulnerability as low, medium, or high, along with some significant criteria. In a follow-up paper~\cite{desmit2017approach}, the authors provided a case study that showed this approach could identify vulnerable node for an independent attack. A CAD file of a jet engine bracket has been modified by the attack exposed the flaw of the assessment. The offense designed with the awareness of the test plan but the testing was unable to detect the attack. 

Tsoutsos et al.~\cite{tsoutsos2017secure} proposed an approach to detect internal structure integrity defects. Their system had two stages. First, they used a compiled tool-path (G-Code) representation to approximate the object which needs to be print. Secondly, they used Finite Element Analysis (FEA) ~\cite{hughes2012finite} to simulate the performance of the printed model under various stress condition. The simulation of FEA returns a color-coded image illustrating any overstressed area. After the observation of the print object and the simulation of the model, manufacturers can determine whether the output model meets the requirement of manufacturing or not.

Turner et al.~\cite{turner2015bad} conducted a study to identify additive manufacturing attack surfaces. Manufacturing toolchains contain several attack vectors that exploit easily. The authors mentioned that generally, there is no secure and authentic way that design files transferred without any types of modification. Design files usually transferred from designers to the manufacturers or users via e-mails or via USB devices. So, they concluded that design files generally incorporate no security measures and thus design files lack integrity. It is also impractical to rely on the quality control process to verify part's integrity as that process is not enough to detect cyber attacks. 

For the FDM printer printed parts, Belter et al.~\cite{belter2015strengthening} presented a paper. They introduced a way named 'fill composition technique' to show how FDM printed manufacturing parts can be strengthening. This method was the first method to counter against the attacks on FDM printed parts. 

In additive manufacturing, the intellectual property might be affected by the act of several adversaries. In 2014, Yampolskiy et al.~\cite{yampolskiy2014intellectual} presented the idea of protection of intellectual properties for the additive layer manufacturing. The paper was first to introduce the intellectual property protection related to layer upon layer manufacturing. The authors examined problems on additive manufacturing, and they introduced a new outsourcing model of additive layer manufacturing-based production.  

For new 3D-printed MEMS magnetometer with optical detection, Khar et al.~\cite{kahr2018novel} proposed a prototyping approach and verification method. Sensitivity measurement with such magnetometer devices indicates high accuracy and excellent structural performance.

\subsection{Legal Aspect of AM Sabotage}

As the AM industry is comparatively new, few people are aware of the legal aspect of AM sabotage. This section surveys those publications which might provide the basis of future explicit discussion of AM sabotage legal aspects.

3D printing services (3DPS) are considered as liability target. Wang et al.~\cite{wang2016breeding} described design defect determination by risk-utility analysis. Consumer expectations, risk, and utility formed his analysis. Major concerns, according to the authors, are whether the 3DPS provide a product without knowing the CAD file details. The authors finished their article by providing information about using the wrong material on the final product, providing legal defense to other actors.

Product liability for recovery theories were proposed by Malloy et al.~\cite{malloy2016three}. While strict liability applies for the commercial sellers and distributors, negligence is applicable for the non-commercial 3D printer uses.  The author's discussion of the topic distinguished between the source of defect and the manifestation of the fault. The nature of additive manufacturing sets the applicability of a particular claim as not as a law. Two instances are the determination of the standards to apply for warnings and instructions and the determination of weather an actor is involved commercially for strict liability to apply. The paper did not discuss the applicability of product liability following a duty to inspect the product that had already been altered by a third party. 

A micro-seller category had been proposed for the "occasional" producer and manufacturer enterprise by Berkowitz et al.~\cite{berkowitz2014strict}. They analyzed the applicability of negligence and breach of warranty and their related defenses. The authors proposed retaining strict liability for 3D printing but creating a new affirmative defenses for the micro-sellers. Berkowitz argued doing so meets the social policies of balancing protection with fairness.

\section{AM framework analysis, authentication and Access control} 

As per our understanding, framework related to the concept of AM first was introduced by Williams et al.~\cite{williams2011functional}. The authors proposed conceptual design for the AM and their functional classifications. The work was the first work as a framework tool for the designers to make the AM technology more precise. 

Based on the previous work, from~\cite{fadhel2013approaches}, Fadhel et al.~\cite{fadhel2014component} proposed a framework which can provide provenance for the 3D printed objects. Information security, transmission, authenticity transferred from the required data with the help of this framework. Data fields include user ID, object ID, time-stamps, user authentication, authorization, etc. They used 3D Object ID (3DOI) to secure the 3D data store and embedded in the object to verify the purpose in the print process against the 3DOI. Then they examine their framework against the attacks on 3D printed objects. The authors considered man-in-the-middle attacks relating to unlicensed copies, illegal access detection, and counterfeit detection. Four proposed systems were available for implementing their framework. The arrangements were RFIDs, digital watermarking, content streaming, and steganography.

In 2014, Mellor et al.~\cite{mellor2014additive} proposed a framework for the implementation of the AM. They suggested the structure along with the review of relevant literature at that time. They have done a case study through interviews, and their framework tested at that time. Their study and implementation work-flow worked as a guide in adopting new and disruptive technology to produce high-value products and generate a high number of products with fewer production faults. 

Go et al.~\cite{go2016framework} also proposed a framework in 2016, which teaches the manufacturers a rapid innovation for the manufacturing process. With the different variety of manufacturing element and manufacturing products, authors mentioned and listed a set of technology in respect to every element by which the manufacturing process can be turned out to be a highly innovative one. 

Frameworks presented by the researchers about the AM technology appeared as a growing and innovative manufacturing technology among the manufacturers. Those frameworks were the first frameworks in AM technology. But, those frameworks didn't provide the logic behind unfortunate implications which can be caused by the intention of the adversaries. For the security reason, a framework which can be considered as manufacturing friendly and against the adversary is vital for the AM.

Yampolskiy et al.~\cite{yampolskiy2016using} proposed a framework to help the researchers to analyze the attacks on AM. From this framework, the researchers have information about diverse attack vectors that could compromise one or more elements of the AM workflow. The compromised factors and their role in AM attacks can determine how adversaries can affect the entire manufacturing process. Adversarial goals might be different for the elements using in the printing process and concerning attack targets. 

Pope et al.~\cite{pope2017hazard} proposed to apply the Systems-Theoretic Process Analysis (STPA) framework ~\cite{young2017system} for the analysis of the hazard in AM. They showed this system used for the identification of manipulations or sabotage attacks. Disruption of timing in the control loop has the hazardous effects; abnormal reading in the sensors like reading too late, too soon or out of order are the other vulnerabilities. They pointed out that disruption and manipulation of power in the 3D printing process can sabotage the print object. 

Gao et al.~\cite{gao2018watching} published a paper related to the safeguarding of the 3D printing process, and they did the safeguarding with real-time perspective. They showed the defense measures against the tool-path or the.STL file attacks and against the attacks related to firmware (i.e., remote firmware attacks). They formulated the attack models and showed defense against the attack models with a very high percentage in result efficiency. Their real-time low-cost pervasive monitoring approach capable of detecting attacks against the cyber-physical attacks on consumer-grade 3D printers. Their defense measures have covered four types of attacks. They included layer thickness attack, infill path attack, printing speed attack, and fan cooling attack.

~\section{Industrial Security Issues and Key Observation}
Additive manufacturing, or known as 3D printing, is a massive plus for the industrial sector in recent time. With the growing need for manufacturing products in every industrial area, it is very much required for the manufacturing industries to supply products with a very high and efficient rate. But, with the cyber-physical and various types of attacks on manufacturing industries, the task to deliver the proper product with effective rate becomes a challenging day by day. In this section, we will enlist the security issues in the industrial sector caused by attacks on the AM.  Then we will provide our crucial observations on the main terms of AM.

\subsection{Industrial Security Issues of AM}
In 2017, Chua et al.~\cite{chua2017standards} presented some basic standard for the 3D printing process, and the authors also presented some criteria for quality control. They also enlisted problems related to industrial production of additive manufacturing. 

Industrial security issues have a significant impact on the manufacturing process. Lipton et al.~\cite{lipton2015additive} first presented the idea of additive manufacturing for the food industry. With the concept of additive manufacturing, health hazard also came in mind of the manufacturers. The inclusion of materials which could be dangerous for the health issue came into consideration at that time. An adversary can affect food production and cause material discrepancy for food products, which could easily affect the human body.  

We discussed several security issues with the production fault and how it could affect the manufacturing product with the stress. First, production fault addressed in 2015 by Huang et al.~\cite{huang2015additive}. They discussed the gaps in industrial sectors and gave potential recommendation for the differences that were present at that time. 

In the medical sector, the AM technology have significant contribution in recent years. Ventola et al.~\cite{ventola2014medical} first presented uses of 3D printing in medical industries and also projected the idea for future users. Gerstle et al.~\cite{gerstle2014plastic} showed the concept of three-dimensional printing in plastic surgery application. Gibson et al.~\cite{gibson2015simplifying} presented the simple idea behind the making of the surgeon into a 3D designer. They provided the idea of how the surgeons can design according to patients need and manufacture 3D-printed objects using the design. Fitzpatrick et al.~\cite{fitzpatrick2016design} presented patient-specific 3D-printed arm cast design. In recent times, researchers are also able to generate workable 3D printed human heart from human tissue. 3D-printing also can be used in neurosurgery. Pucci et al.~\cite{pucci2017three} presented the idea of 3D-printing in neurosurgery and by far extent, they provided limitations also. Limitations of 3D-printing or AM in the medical sector is severe rather than mechanical limitations. For the limitations of 3D-printing or additive manufacturing in medical sectors, a huge number of patients life can endanger if the application of the 3D-printing is not correct. In recent time, doctors tend to prescribe 3D-printed artificial legs for the persons who might have lost their leg for accident or other reason. But, in some cases, it is common that, those 3D-printed medical applications are not able to provide the best for the patients. Some of them lack the usefulness of a printed part as a whole.

\subsection{Our Key Observations}
From this survey, we have enlisted and discussed major potential flaws in the manufacturing process of additive manufacturing. From the study and discussion, we have come with some critical observations on AM and manufacturing process in terms of security and efficient manufacturing. 

\begin{framed}
First observation: The machine capabilities for different manufacturing products varies for manufacturing. Manufacturing materials have a more substantial impact on manufacturing products because they set the main bar for AM production. There are also some quality benchmarks setting the bar for the AM products to be reasonable to produce. 
\end{framed}

\begin{framed}
Second observation: A general and secure framework for authentication and access control has not been implemented yet for all AM devices and manufacturing process. Lack of secure authentication creates a problem relating to identity theft and theft of technical data. 
\end{framed}

\begin{framed}
Third observation: AM devices should not be on the public network with weaker authentication security. An adversary can quickly introduce malicious intent in the time of manufacturing. Theft of identical and technical data also are reasons for not recommending the public network for the AM process.  
\end{framed}

\begin{framed}
Fourth observation: Proper choice of print materials provides a better and stable manufacturing product with higher durability. 3D print materials are very distinct and have separate durability issues. The manufacturer should use proper filament for the FDM-based printing process, which eventually provides higher durability and low-risk factor for the products. Using print materials wisely also ensures low-cost printing process. 
\end{framed}

\begin{framed}
Fifth observation: The problem which caused vulnerabilities of different types of 3D-printer software are quite similar because 3D-printer software has only the desktop version, which lacks the secure authentication process. Cloud based AM services also lacks authentication and vulnerable to cyber attacks. Local networks should not be used or need to be secure against an adversary for the defense against the attacks on 3D printing.  
\end{framed}

\section{Conclusion}

In this survey, we presented the constraint, security, and privacy issues in additive manufacturing (AM). We showed the limitations of the AM process, and then we gave the probable solution of the restrictions. We also studied the classifications of attacks on AM, and then we listed the defense measures for the respective offenses. Later, we analyzed AM frameworks to give the readers proper ideas to have secure authentication and access control in the AM process. Finally, we presented briefly about the security issues in industrial sectors and our crucial observation from the survey. 

Overall, the safety of the AM process and manufacturing product today depends on frameworks, technologies, and security mechanisms implemented by each manufacturing individual. Based on the specific cases, AM process can be vulnerable for several types of attacks. Those attacks indicate the urgency of creating a secure and general framework for the manufacturer. AM manufacturer needs to work closely with supervisory agencies and standardization organizations to tackle newly emerged threats and to develop a robust and secure standard for the AM process.

\section*{Acknowledgment}
This work is supported in part by US NSF under grants CNS-1812553. Shengmin Guo is supported by NSF-Consortium for innovation in manufacturing and materials (CIMM) program (grant number $\#$ OIA-1541079).


%





\ifCLASSOPTIONcaptionsoff
  \newpage
\fi





\bibliographystyle{IEEEtran}
\bibliography{IEEEabrv,Bibliography}
\vspace{-7 mm}	

\begin{IEEEbiography}[{\includegraphics[width=1in,height=1.25in,clip,keepaspectratio]{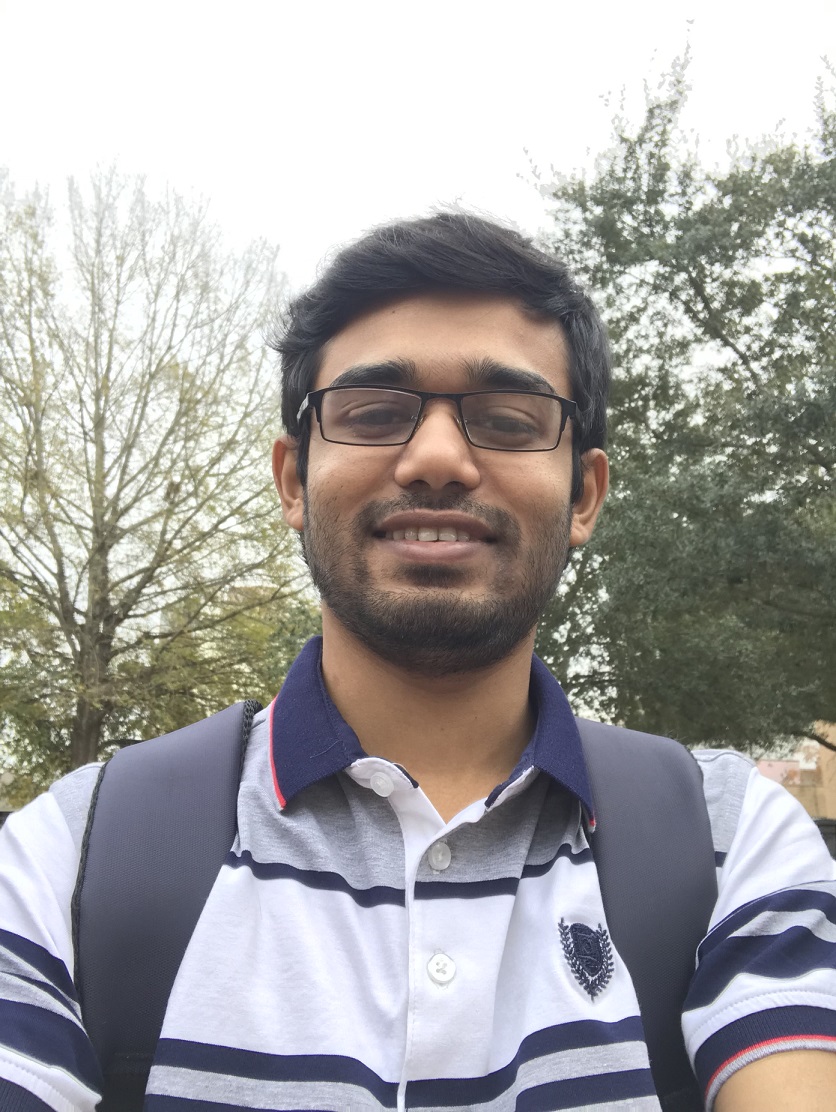}}]{Md Nazmul Islam} received a B.Sc degree in Computer Science $\&$ Engineering from Rajshahi University of Engineering and Technology (RUET), Rajshahi, Bangladesh, in 2018. He is currently a Ph.D. student in the School of Computing and Informatics at the University of Louisiana at Lafayette. His research interests include additive manufacturing, security, and privacy issues of embedded devices.
\end{IEEEbiography}
\vspace{-7 mm}	

\begin{IEEEbiography}[{\includegraphics[width=1in,height=1.25in,clip,keepaspectratio]{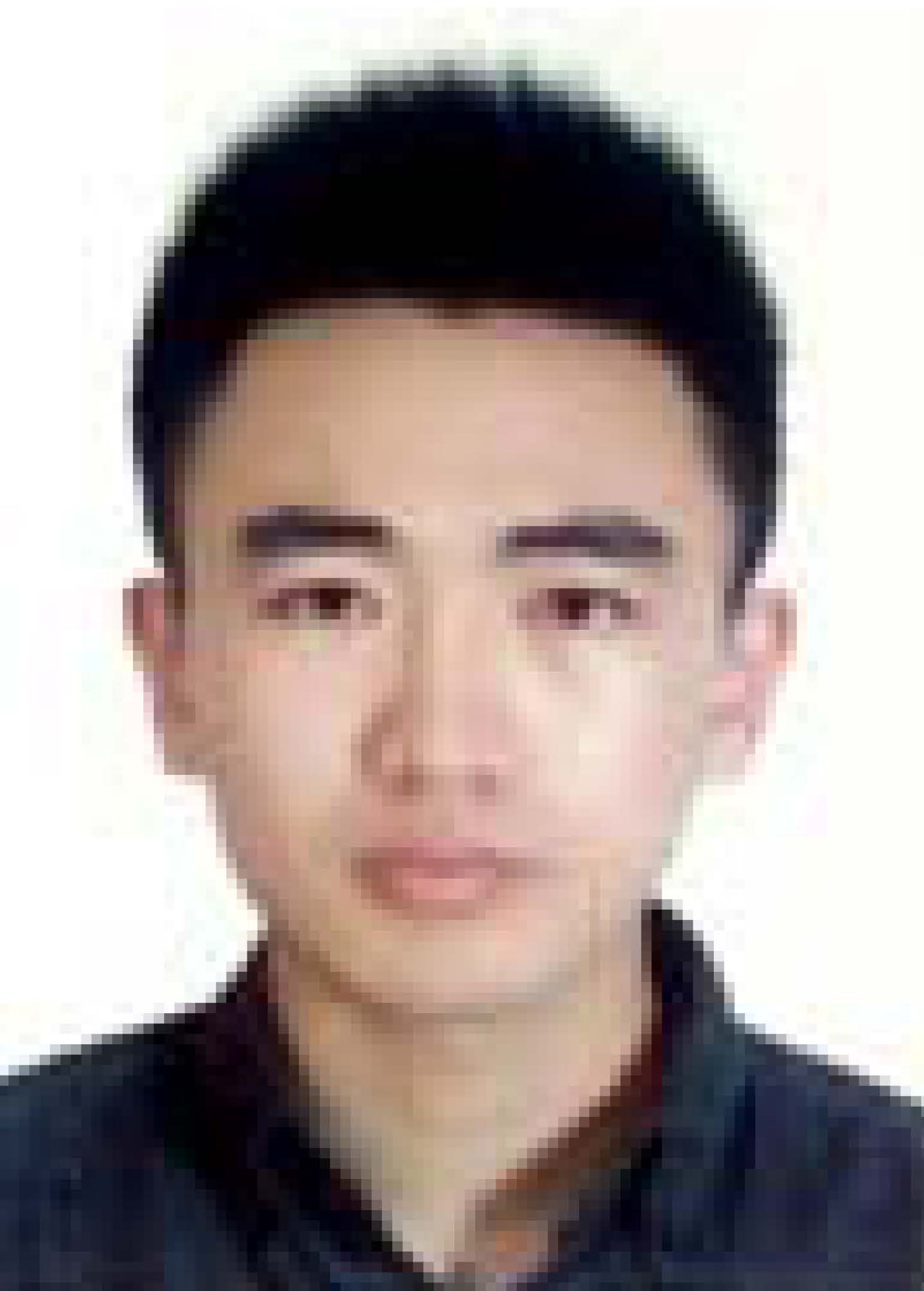}}]{Yazhou Tu} received the B.S. degree in software engineering from Wuhan University, Wuhan, China, in 2013, and the M.S. degree in software engineering from Tsinghua University, Beijing, China, in 2016. He is currently persuing the Ph.D. degree in the School of Computing and Informatics at the University of Louisiana at Lafayette. His research interests include security and privacy of sensors, security of embedded systems, and protocol design.
\end{IEEEbiography}
\vspace{-7 mm}	

\begin{IEEEbiography}[{\includegraphics[width=1in,height=1.25in,clip,keepaspectratio]{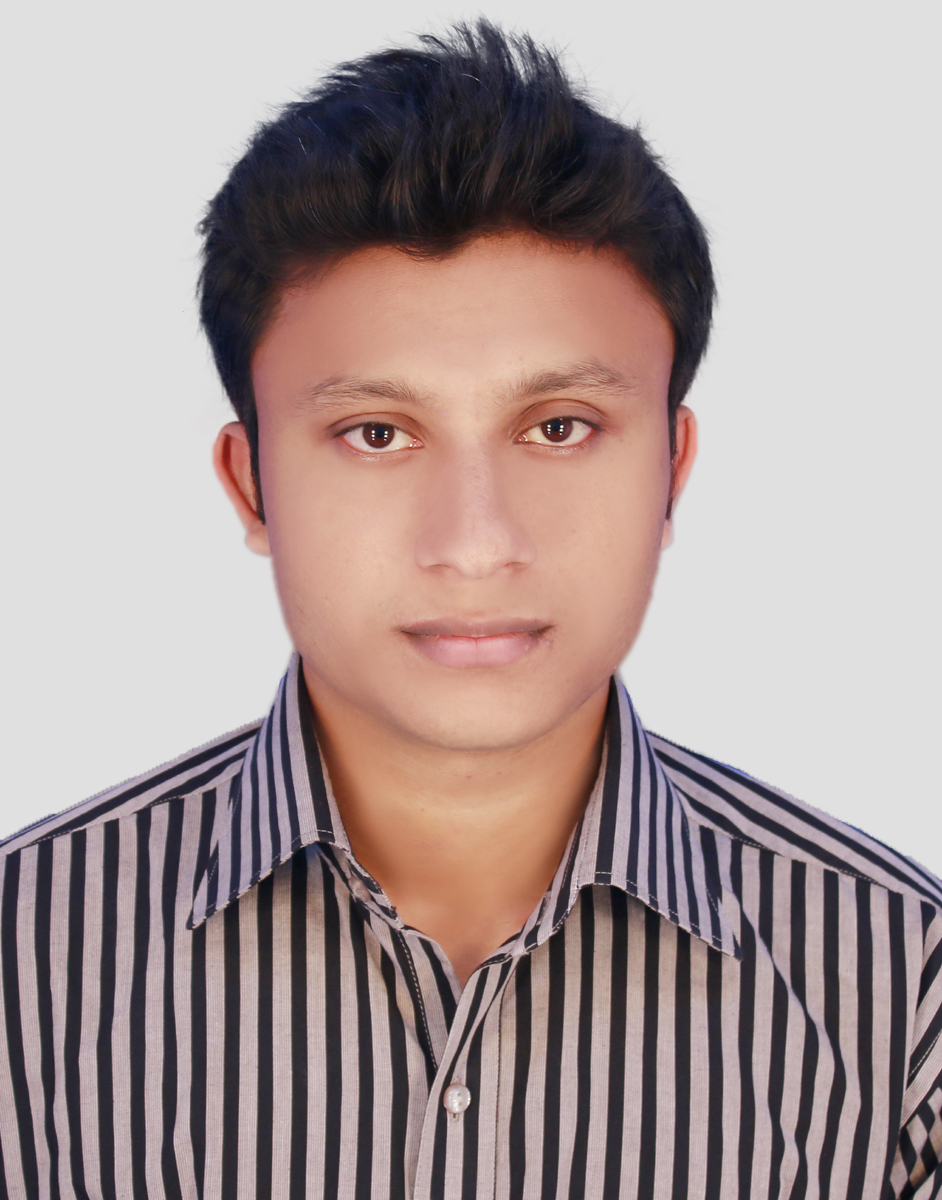}}]{Md Imran Hossen} received a B.S. degree in Electronics and Communication Engineering from Khulna University of Engineering $\&$ and Technology (KUET), Khulna, Bangladesh, in 2016. He is currently a Ph.D. student in the School of Computing and Informatics at the University of Louisiana at Lafayette. His research interests lie primarily in the areas of web security and cybersecurity. He is also interested in machine learning for offensive security.
\end{IEEEbiography}
\vspace{-7 mm}	

\begin{IEEEbiography}[{\includegraphics[width=1in,height=1.25in,clip,keepaspectratio]{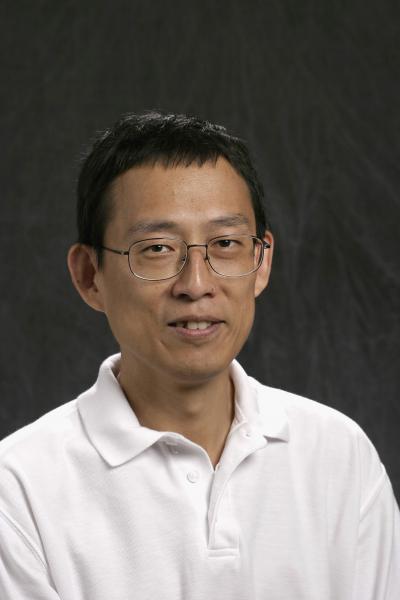}}]{Shengmin Guo} is a professor in Department of Mechanical and Industrial Engineering and the Director of Turbine Innovation and Energy Research (TIER) center at Louisiana State University. He is also a Louisiana registered professional engineer. Dr. Guo was a lecturer at the University of Oxford (England) and the University of Manchester Institute of Science and Technology (England) from 1998 to 2004. He joined LSU Mechanical Engineering Department in August 2004 as an assistant professor and was promoted to professor in 2014. Dr. Guo's research activities are in the fields of thermal fluids, instrumentation, power generation, laser additive manufacturing, and high temperature materials. Dr. Guo has established a track-record on gas turbine aerodynamics and heat transfer, fuel cells, plasma spray coatings, high temperature ceramics and alloys, and advanced manufacturing.  His research projects are funded by NSF, NASA, DOE, Air Force, LaSPACE, Louisiana Board of Regents, and industry. His current research focus is on laser based additive manufacturing.
\end{IEEEbiography}
\vspace{-7 mm}	

\begin{IEEEbiography}[{\includegraphics[width=1in,height=1.25in,clip,keepaspectratio]{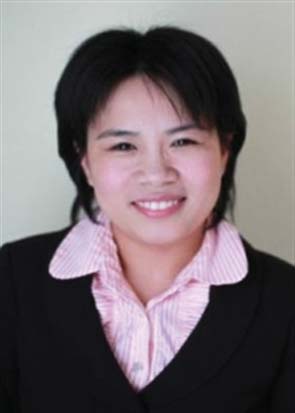}}]{Xiali Hei} received the B.S. degree in electrical engineering from Xi'an Jiaotong University, Xi'an, China, in 2002, the M.S. degree in software engineering from Tsinghua University, Beijing, China, in 2005, and the Ph.D. degree in computer science from Temple University in 2014. She is an assistant professor in the School of Computing and Informatics at the University of Louisiana at Lafayette. Prior to joining the University of Louisiana at Lafayette, she was an assistant professor at Delaware State University from 2015-2017 and Frostburg State University 2014-2015. Her research interests are secure real-time wireless medical devices, cyber-physical attacks, vulnerability assessment and malware detection on Android, and efficient encryption schemes design. She was awarded NSF CRII grant and Delaware DEDO grant. She got several awards such as: ACM 2014 MobiHoc Best Poster Runner-up Award, Dissertation Completion Fellowship, The Bronze Award Best Graduate Project in Future of Computing Competition, IEEE INFOCOM and IEEE GLOBECOM student travel grant, etc. Her papers were published at USENIX Security Symp., ACM CCS, IEEE INFOCOM, etc. She is the TPC member of USENIX Security, IEEE GLOBECOM, IEEE ICC, WASA, etc. 
\end{IEEEbiography}

\vspace{-7 mm}

\vfill


\end{document}